\documentclass[preprint, amsmath, amssymb, preprintnumbers, showpacs, showkeys, aps, prl]{revtex4-1}
\usepackage{graphicx}
\usepackage{float}
\usepackage{color}
\begin{document}
\preprint{}

\title{Quasi One Dimensional Dirac Electrons on the Surface of Ru$_2$Sn$_3$}

\author{Q.D.Gibson$^{1}$}
\author{D. Evtushinsky$^{2}$}
\author{A.N.Yaresko$^{3}$}
\author{V.B. Zabolotnyy$^{2}$}
\author{M.N.Ali$^{1}$}
\author{M.K.Fuccillo$^{1}$}
\author{J. Van den Brink$^{2}$}
\author{B. B{\"u}chner$^{2}$}
\author{R.J.Cava$^{1}$}
\author{S.V.Borisenko$^{2}$}

\affiliation{$^1$Department of Chemistry, Princeton University, Princeton, New Jersey 08544, USA}
\affiliation{$^2$Leibniz Institute for Solid State and Materials Research, IFW Dresden,
D-01069 Dresden, Germany}
\affiliation{$^3$Max-Planck-Institut fur Festkorperforschung Heisenberggstrabe 1, 70569 Stuttgart, Germany}
\date{\today}


\begin{abstract}

We present an ARPES study of the surface states of Ru$_2$Sn$_3$, a new type of a strong 3D topological insulator (TI). In contrast to currently known 3D TIs, which display two-dimensional Dirac cones with linear isotropic dispersions crossing through one point in the surface Brillouin Zone (SBZ), the surface states on Ru$_2$Sn$_3$ are highly anisotropic, displaying an almost flat dispersion along certain high-symmetry directions. This results in quasi-one dimensional (1D) Dirac electronic states throughout the SBZ that we argue are inherited from features in the bulk electronic structure of Ru$_2$Sn$_3$, where the bulk conduction bands are highly anisotropic. Unlike previous experimentally characterized TIs, the topological surface states of Ru$_2$Sn$_3$ are the result of a d-p band inversion rather than an s-p band inversion. The observed surface states are the topological equivalent to a single 2D Dirac cone at the surface Brillouin zone.\end{abstract}

\maketitle

Electrons in condensed matter generally behave as semi-classical particles. In certain materials systems, however, electrons behave relativistically and thus are referred to as Dirac electrons. Dirac electrons found in two dimensions in graphene\cite{novoselov2005two} and on the surfaces of topological insulators (TIs), as well as in a small number of three-dimensional bulk systems (i.e. Cd$_3$As$_2$, Na$_3$Bi, PbSnSe, and Bi)\cite{hsieh2012topological, liang2013evidence, wang2012dirac, liu2013discovery, borisenko2013experimental, neupane2013observation, wang2013three} The Dirac electron systems that are characterized to date are nearly isotropic, that is, the electrons' Fermi velocities are similar for all directions of propagation. The electronic structure thus forms what is known as a "Dirac Cone". Recently, anisotropic Dirac cones have been predicted for Ag$_2$Te and HgS, and also on the side surfaces of weak topological insulators \cite{Ag2Te,virot2011metacinnabar, yan2012prediction} but experimental observations of these states have not yet been reported. Here we report the observation of highly anisotropic massless Dirac states on the surface of Ru$_2$Sn$_3$, with energies within the bulk band gap. The surface states disperse linearly with a high velocity in one direction on the surface, but are nearly non-dispersive in the perpendicular direction, giving rise to a quasi 1D massless Dirac state and a very unusual surface Fermi surface. Unlike previously characterized topological insulator systems, the band inversion that generates the exotic surface state in Ru$_2$Sn$_3$ is due to d-p band inversion, not s-p band inversion. The quasi 1D Dirac states observed here open up a new avenue for studying the properties of relativistic electrons in condensed matter systems.

The crystal structure of the low temperature orthorhombic phase of Ru$_2$Sn$_3$ is shown in Fig 1(a), with an emphasis on the quasi 1D chains formed by Ru(2) and Sn atoms along the c-axis. The transport properties of Ru$_2$Sn$_3$ are shown in Figure 1(b). The resistivity (large panel) shows an unusual trend, the broad decrease in resistivity at intermediate temperatures likely being a reflection of the broad tetragonal to orthorhombic phase transition that starts at 200K \cite{susz1980diffusionless, poutcharovsky1975diffusionless}. The magnitude of the resistivity and its temperature dependence are typical of those seen for high carrier concentration small band gap semiconductors. An n-p crossover is also seen in the temperature dependence of the Seebeck coefficient (upper inset) at about 200K. The magnitude of the low temperature Seebeck coefficient (-80$\frac{\mu V}{K}$ at 50K) is very large, consistent with that of a doped small band gap semiconductor\cite{noguchi2007low}.

This quasi 1D differentiation of the Ru atoms into chains is apparent in the calculated electronic structure of the bulk orthorhombic phase (Fig 1(c)). The calculations confirm the expected gapped nature of Ru$_2$Sn$_3$ at 14 electrons per Ru\cite{fredrickson2004nowotny2}. Most of the Ru 4d orbitals are highly dispersive in energy and strongly hybridized with the Sn 5p states. The calculations show that without spin orbit coupling (SOC) (not shown), there is a direct overlap between the Ru 4d based conduction band and the Sn 5p based valence band at $\Gamma$. Inclusion of SOC opens a gap, however, leaving behind an inverted band structure. This is a d-p band inversion between Sn 4p orbitals and the Ru 4d$_{z^{2}}$ and 4d$_{x^{2}-y^{2}}$ orbitals of one of the independent Ru atoms (Ru(2)) in the structure. These Ru orbitals, in contrast with the other Ru 4d states, are highly anisotropic, they are not strongly hybridized with the Sn states and are weakly dispersive in the k$_z$=0 plane, especially in the $\Gamma$-X and $\Gamma$-S directions. They do disperse in the k$_z$ direction and along $\Gamma$-U, likely due to the presence of the quasi 1D Ru(2)-Sn chains.  This interaction of fully dispersive 5p states and strongly anisotropic 4d states forms the electronic structure of the bulk phase near the Fermi level, and therefore is expected to dominate the physics and contribute most to surface states formed within the band gap. This type of electronic structure is in strong contrast with currently known TIs, which exhibit three dimensional bulk valence and conduction bands that strongly disperse in all directions. In addition, the band inversion is caused by the effect of SOC on d-p overlap in Ru$_2$Sn$_3$, in contrast to the case in previous TIs, where the band inversions have an origin in s-p overlap. That there are the usual precision issues with DFT calculations near the band gap of Ru$_2$Sn$_3$ is evidenced by the fact that the calculated electronic structure shows a small indirect band overlap leading to a semi-metal state, in contrast to transport measurements and ARPES data, which clearly indicates that Ru$_2$Sn$_3$ has a full band gap (Fig 1(e)). Thus though at this level of theory the calculated inverted bands have the same inversion eigenvalue and a therefore a predicted trivial band inversion, the presence of many conduction bands with different parities very close in energy near E$_F$ leads to ambiguities that future in-depth calculations and theoretical analysis will have to resolve.    

The ARPES measurements clearly show presence of the bulk valence band with a dispersion consistent with the electronic structure calculations, and no indication at any photon energy of the bottom of the conduction band. Thus the Fermi energy in the as-prepared Ru$_2$Sn$_3$ crystals falls clearly within the bulk band gap. Surface electronic states are clearly observed. These surface states display an unusual "star" shaped Fermi surface (Fig 2(a)), which is very different from the nearly circular momentum distributions for the surface Fermi surfaces observed in previous TIs\cite{xia2009observation}. The lines connecting the "stars" appear on casual inspection to be of hexagonal or psuedo-hexagonal symmetry; the symmetry of the crystal is orthorhombic, however, and the pseudo-symmetry is simply that of the cleavage plane. The fact that these are indeed surface states is seen in Figure 2(b), which shows the valence band spectrum at multiple photon energies. The observed states that make up the Fermi surface do not qualitatively change with photon energy, meaning that they display no dispersion perpendicular to the cleavage surface and are therefore surface states. In contrast, the bulk valence band (BVB)  changes qualitatively upon changing photon energy, allowing for identification of separate bulk and surface states and the identification of the bulk band gap. A closer look near E$_F$ (Fig 2(c)) shows the surface states originating from the BVB (the high intensity features in the spectra), then separating from it, and crossing E$_F$. This indicates that these are metallic surface states. A clearer look at these surface states (Fig 2(d)) reveals that they are linearly dispersing bands, typical of what is seen for a Dirac dispersion.

A schematic of the features visible in the surface Fermi surface map is shown in Figure 3(a). To understand the underlying surface electronic structure leading to such an anisotropic intensity distribution at the Fermi level, more than 50 line cuts were taken perpendicular to the $\bar{\Gamma}$-$\bar{X}$-$\bar{\Gamma}s$ line in momentum space (Fig. 3a) in a relatively narrow energy interval (250 meV) near E$_F$. Several exemplary datasets from these cuts are shown in Fig. 3(b). It is clearly seen that two linear dispersions cross the Fermi level close to $\bar{\Gamma}$; these result in the two rays of the "star" seen on the surface Fermi surface. The two linear dispersions converge to a single crossing of E$_F$ on approaching the $\bar{X}$ point, and then diverge again. In the vicinity of the surface X point itself, their crossing point is below the Fermi level. For a quantitative analysis, we fitted the momentum distribution curves (MDCs, a typical example is shown in Fig. 4(c)) with two Lorentzians, and show the results as red curves in Fig. 3(c). (At the $\bar{X}$ point, a single Lorentzian was employed, see below.) We fit the obtained dispersions away from the $\bar{X}$ point with straight lines, and thus determined the energy position of each crossing. These crossings create the quasi 1D "Dirac lines". A selection of the corresponding energy values are given in Fig. 3 (c); they confirm the situation qualitatively implied by the data in Fig. 3(b) - the crossing points of the linear dispersions continuously shift away from the Fermi level, from being above E$_F$ at Gamma to below E$_F$ near $\bar{X}$.

In order to determine whether the linear bands cross, or whether there is a gap in the surface states that implies a finite mass for these quasi 1D Dirac electrons, the spectra near $\bar{X}$ were closely analyzed. These spectra can be described by a single Lorentzian in the energy region between the crossing and Fermi level. This feature has practically infinite Fermi velocity and results in the narrow lines on the Fermi surface maps. This situation closely resembles what is seen for the substrate-induced band gap on graphene\cite{zhou2007substrate}, where the gap region can extend up to 0.5 eV and is characterized by a single peak in the momentum distribution with no evidence for the existence of two features dispersing away from each other. In addition, the momentum width of the line connecting the rays of the star is not constant: it has two local minima straddling the $\bar{X}$-point. This asymmetry of the width near the minimum is a typical sign of the opening of the gap and the acquisition of mass. These are the points where the flat dispersions going from $\Gamma$ to X cross the Fermi level. We therefore conclude that there is a gap in the vicinity of the $\bar{X}$-points with the spectral weight at the Fermi level explained by the in-gap intensity as in \cite{zhou2007substrate}. The data therefore show that the surface states in Ru$_2$Sn$_3$ can be described by the electronic structure shown in Fig. 3(d); the implication is that there is a single Dirac point around 150 meV above E$_F$ at $\bar{\Gamma}$ that originates from the band inversion around $\Gamma$ in the bulk.

Our data imply that Ru$_2$Sn$_3$ is a strong TI with a surface Dirac point at $\Gamma$, in which a very small mass is induced along the $\bar{\Gamma}$-$\bar{X}$-$\bar{\Gamma}$ line, creating a quasi 1D topological surface state. We propose that this strong anisotropy is induced by the quasi 1D character of the Ru 4d orbitals in the bulk conduction band. To support the interpretation of these as topological surface states, a linear fit to the dispersion of the momentum distribution curves (MDC) over a much broader energy range is shown in Figure 4(a). The state is highly linear, consistent with the Dirac spectrum and the surface states of known TIs. The broadening of the states, shown in the Figure 4(b) is also consistent with well-defined topologically protected surface states \cite{ScatteringTopo}, with the broadening only becoming apparent when the surface feature merges with the bulk valence band. A typical MDC and Energy distribution curve (EDC) is shown in Figure 4(bc. Finally, a schematic of the surface states observed is shown in figure 4(d). While the surface states on Ru$_2$Sn$_3$ are highly unusual, they can be continuously deformed into that of the usual Dirac cone, meaning that the two are topologically equivalent. Our results show that it is possible for strong TIs to exhibit surface states that are almost 1D in nature, with massless Dirac fermions at the zone center and massive Dirac fermions (with a very small mass) at the zone edges. While these states are topologically equivalent to those observed in previous TIs, they are expected to behave very differently in transport experiments. Their study may open up a new avenue and experimental platform for determining the effects of dimensionality and anisotropy in relativistic Dirac fermion systems.

\section{Methods}
Ru$_2$Sn$_3$ single crystals were grown from a mixture of Ru and Sn in a Bi flux, with the stoichiometry Ru$_2$Sn$_3$Bi$_{67}$. Phase identity was confirmed by X-ray diffraction (XRD), as was the previously reported orthorhombic crystal structure at 100 K \cite{susz1980diffusionless, poutcharovsky1975diffusionless} for which single crystal data was refined at 100K (for the refined values, see supplementary table S1). The cleavage plane of single crystals was determined by XRD to be the 113 plane of the room temperature tetragonal phase.

ARPES data were collected at Helmholtz Zentrum Berlin using the synchrotron radiation from the BESSY storage ring. Excitation photon energies in the range 15 eV to 110 eV were used under full control of polarization. The end-station 1$^{3}$-ARPES is equipped with a 3He cryostat that allows one to keep the temperature of the sample at ~1K. All spectra in the present work were recorded at such temperatures. The overall energy resolution depends on the photon energy and here was set to about 3 meV at 15 eV and about 12 meV at 110 eV. Momentum resolutions along (k$_y$) and perpendicular (k$_x$) to the slit are ~0.2 degrees. All samples were cleaved at approximately 40 K, exposing perfectly flat surfaces corresponding to the 113 cleavage plane in the tetragonal coordinates.

Electronic structure calculations were performed by density functional theory (DFT) using the Wien2k code with a full-potential linearized augmented plane-wave and local orbitals basis, together with the Perdew-Burke-Ernzerhof parameterization of the generalized gradient approximation\cite{perdew1996generalized}.The plane wave cutoff parameter RMTK$_{max}$ was set to 7 and the Brillouin zone (BZ) was sampled by 2000 k-points. For all calculations, the experimentally determined orthorhombic structure at 100K was used. The contribution of atomic orbitals to the bands were determined through the use of "fat band" plots, which show the contribution of a specific atomic orbital to each state along the selected k-path. Qualitatively similar results were also obtained using the fully relativistic PY LMTO code.

The temperature dependent resistivity measurements were carried out in a Quantum Design physical property measurement system (PPMS). Thermopower measurements were performed using a homemade probe in conjunction with an MMR Technologies SB100 Seebeck measurement system. Resistivity and thermopower measurements were taken on single crystal samples. 

\section{Acknowledgements}

The research at Princeton was supported by the ARO Sponsored MURI on superconductivity, grant W911NF-12-1-0461. The research in
Dresden was supported by the DFG grants BO1912/3-1, BO1912/2-2 and ZA.
654/1-1.

\bibliography{Ru2Sn3_paper_bibliography}

\begin{figure}[htbp]
\begin{center}
\includegraphics[scale=0.65]{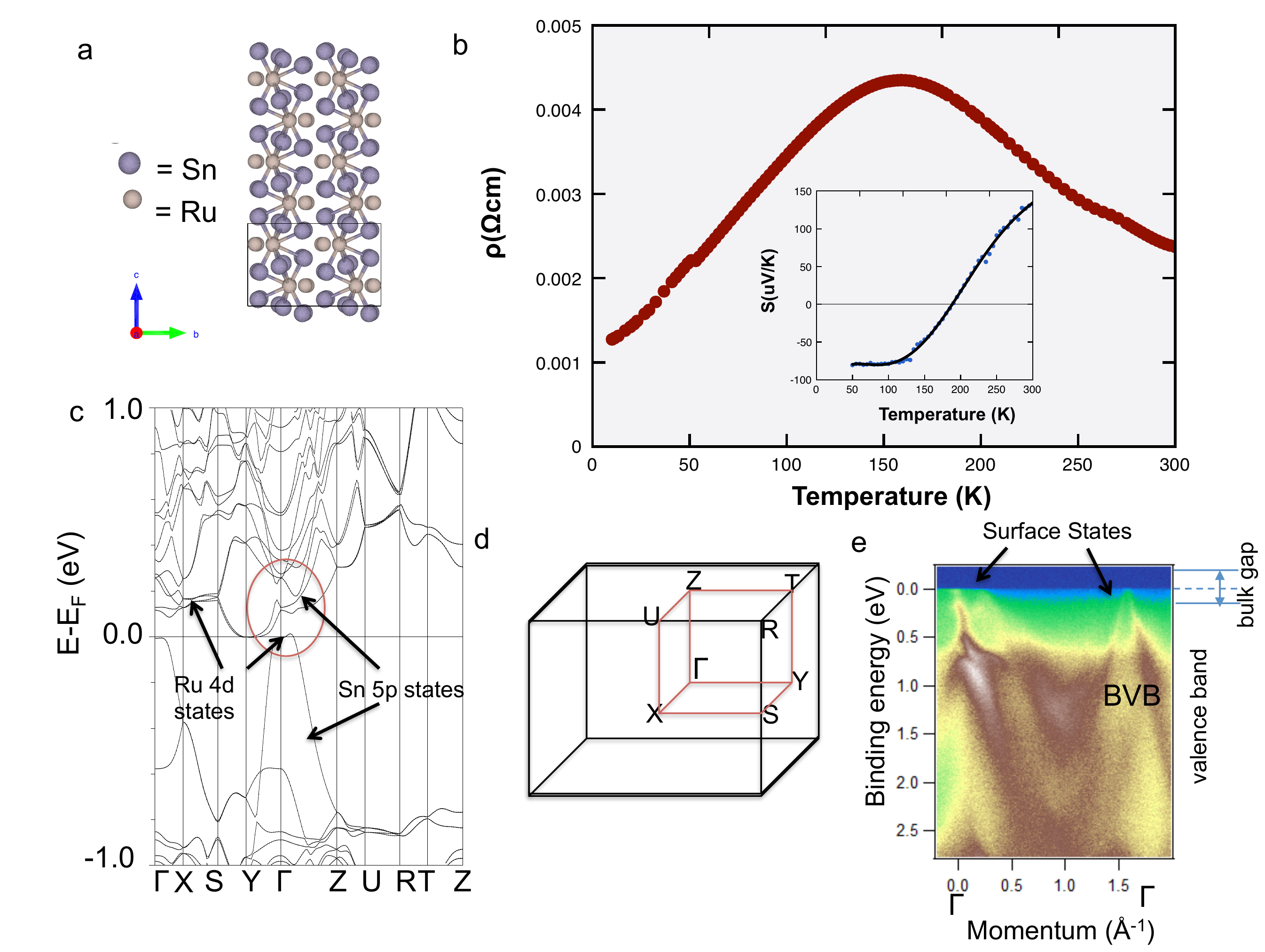}
\caption{(a) Crystal structure of Ru$_2$Sn$_3$ with emphasis of the formation of the Ru-Sn chains that form the electronic structure near E$_F$. The closest Ru(2)-Sn bonds are drawn (b) Temperature dependence of the resistivity (main panel) and Thermopower (inset). A line is drawn as a guide to the eye. (c) Calculated electronic structure of Ru$_2$Sn$_3$, with Ru 4d$_{x^{2}-y^{2}}$,  Ru 4d$_{z^{2}}$  and Sn 5p states highlighted. The red circle shows the area over which the p-d band inversion takes place.(d) Brilloun zone of Ru$_2$Sn$_3$ (e) ARPES spectrum showing the valence band and surface states, indicating the presence of a bulk gap.}
\label{default}
\end{center}
\end{figure}

\begin{figure}[htbp]
\begin{center}
\includegraphics[scale=0.65]{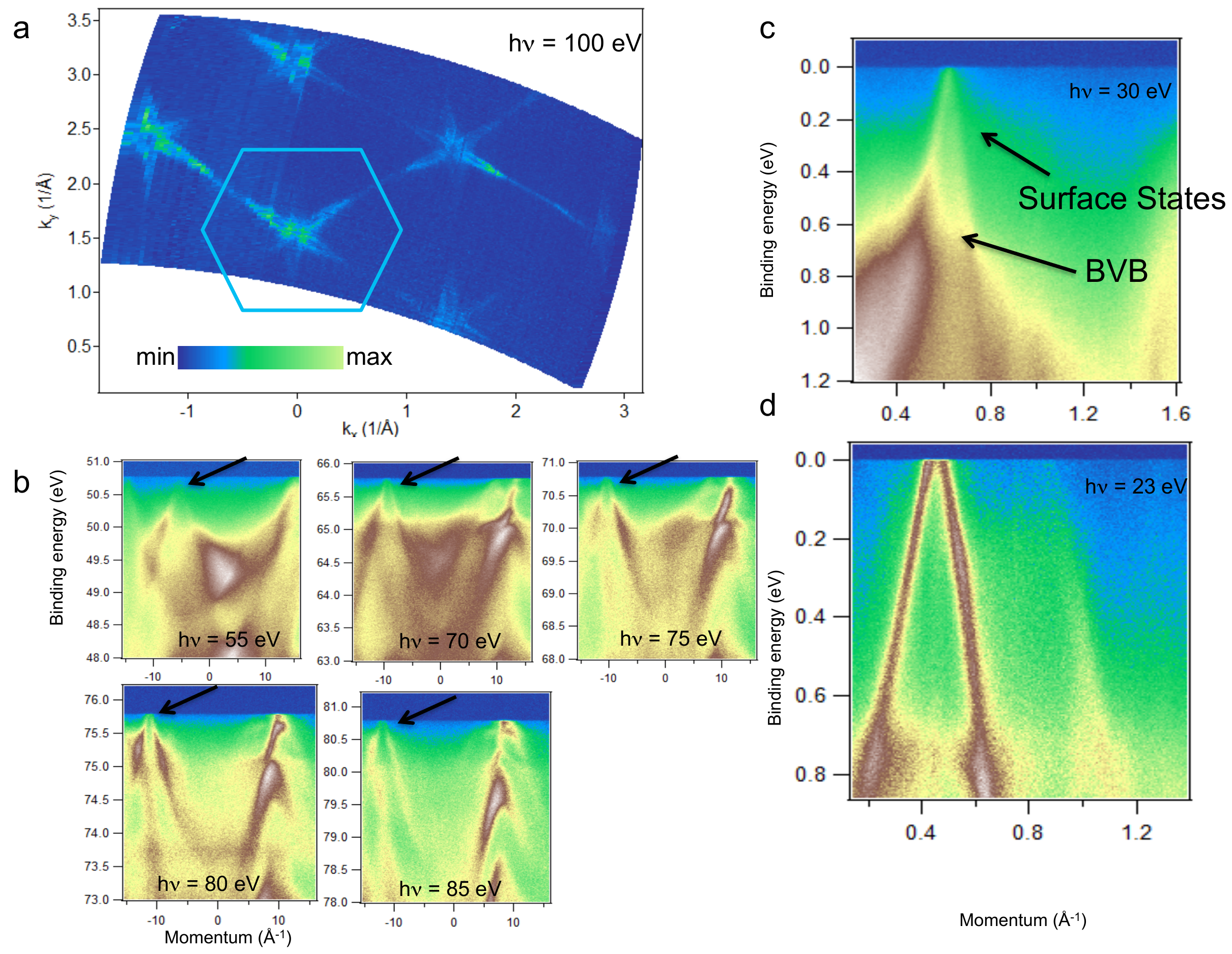}
\caption{(a) Fermi surface of Ru$_2$Sn$_3$ comprised of surface states. The surface Brillouin zone is drawn in blue. (b) Momentum energy cuts taken at different photon energies showing the bulk valence band and surface states. The bulk valence band changes qualitatively with photon energy while the surface states that come to E$_F$ do not. (c) Momentum-energy cut showing the surface states originating from the bulk valence band and coming to E$_F$. (d) detail of a momentum-energy cut of the surface state, showing the linear dispersion.}
\label{default}
\end{center}
\end{figure}

\begin{figure}[htbp]
\begin{center}
\includegraphics[scale=0.65]{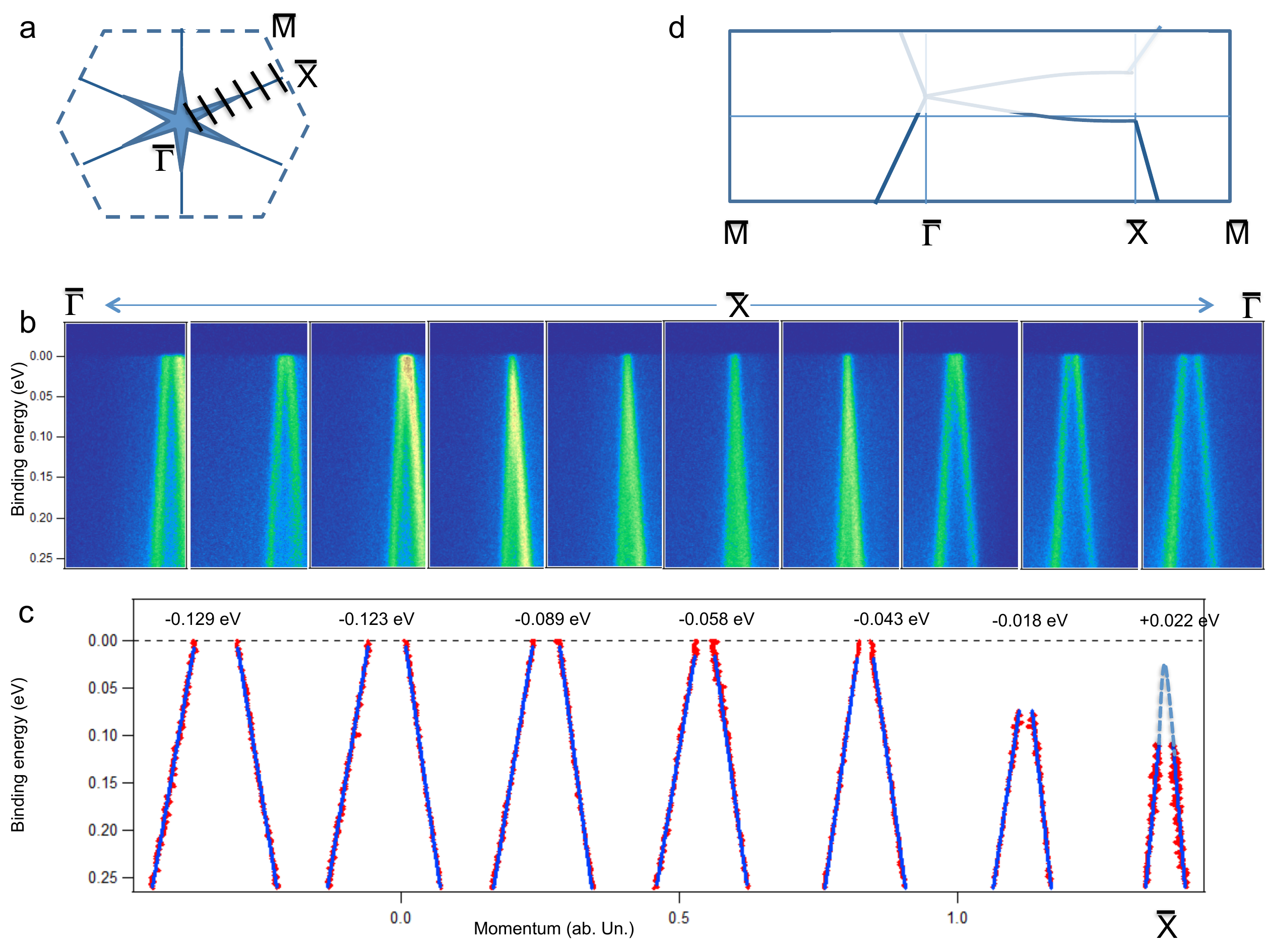}
\caption{(a) Cartoon schematic of the Fermi surface, with black lines showing a schematic of the cuts shown in (b). (b) Momentum energy cuts going through different points ranging from through $\bar{\Gamma}$ to $\bar{X}$ (left to right) showing linearly dispersing states with slightly changing Dirac point energies.(c) Linear fits to the data approaching X, showing the extrapolated Dirac point energy at the top (or where the linear bands meet). Approaching X the states become blurry due to gap formation; further lines are added as a guide at the X point.(d) Schematic of the surface state electronic structure outlining the most likely scenario. Observed bands are shown as darker.}
\label{default}
\end{center}
\end{figure}

\begin{figure}[htbp]
\begin{center}
\includegraphics[scale=0.65]{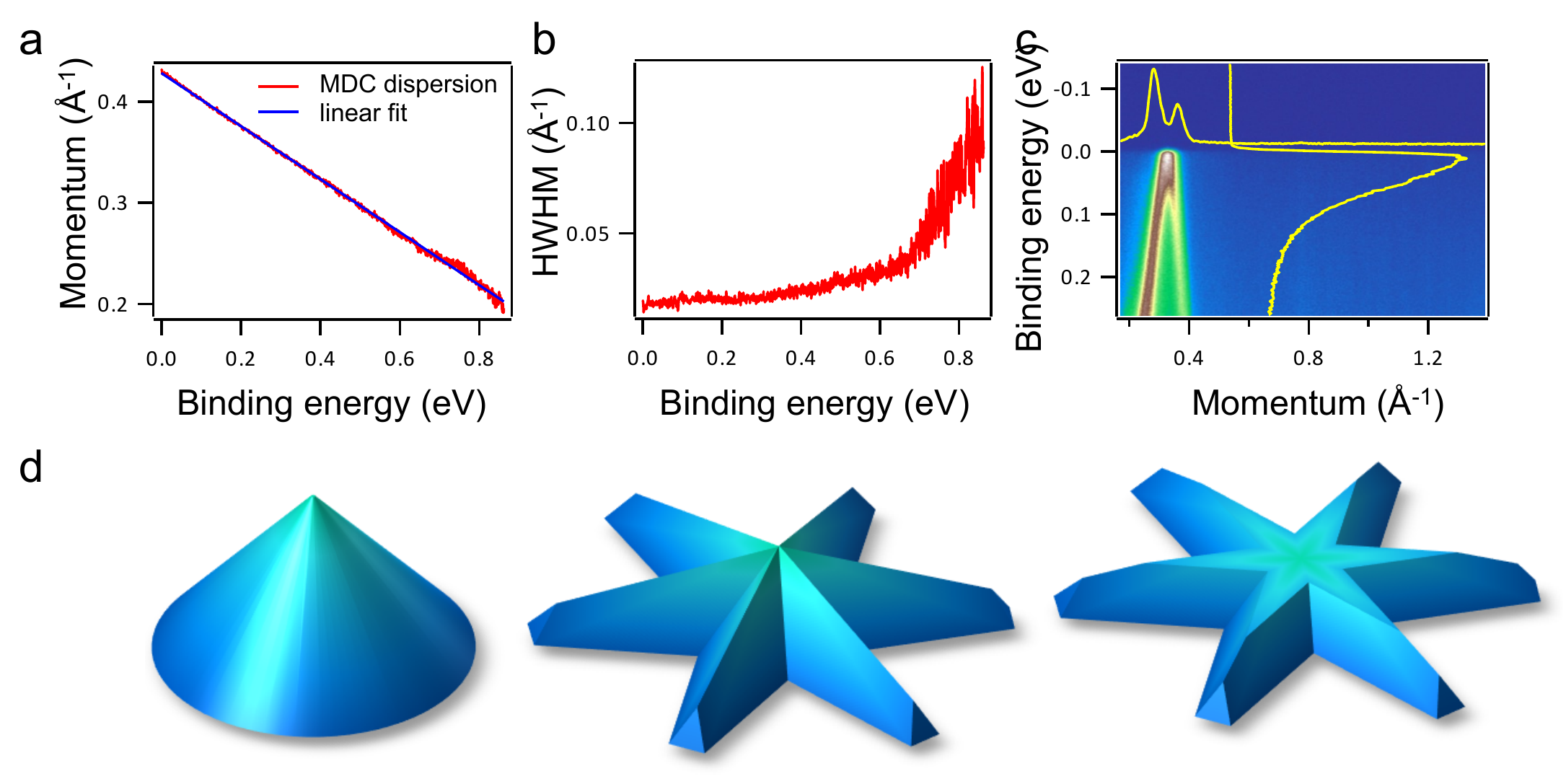}
\caption{(a) Linear fit to the MDC dispersion showing the highly linear dispersion. (b) Half width half maximum vs. energy showing very low broadening of the surface state.  (c) Example MDC and EDC of the surface state (d) Cartoon schematic showing (left to right) the typical Dirac cone, the topologically equivalent highly anisotropic Dirac cone in Ru$_2$Sn$_3$ and the states observed by ARPES}
\label{default}
\end{center}
\end{figure}

\begin{figure}[htbp]
\begin{center}
\includegraphics[scale=.9]{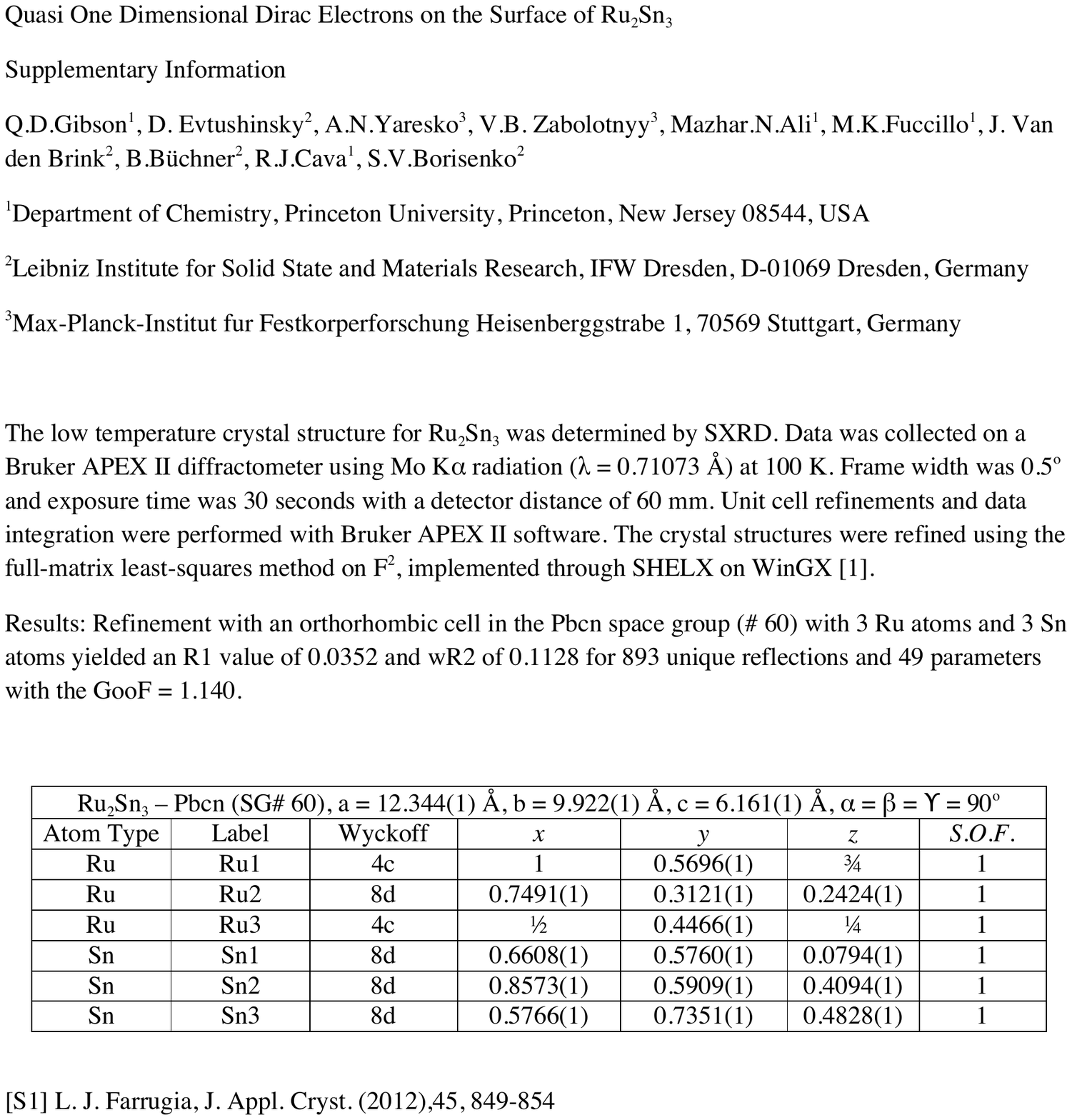}
\label{default}
\end{center}
\end{figure}
\end{document}